\documentclass[a4paper]{article}
\usepackage{ISCSLP2026}
\usepackage{ifthen}
\usepackage{multirow}
\usepackage[hidelinks]{hyperref}

\newboolean{blind}
\setboolean{blind}{false} 

\makeatletter
\def\@oddhead{}
\def\@evenhead{}
\makeatother

\title{Steering Where to Listen: Instruction-Based Activation Steering Redirects Temporal Attention in Large Audio-Language Models}

\name{
	\ifthenelse{\boolean{blind}}{Anonymous to ISCSLP}
	{Tsung-En Lin$^1$, Hung-Yi Lee$^{1,2}$}
}

\address{
  \ifthenelse{\boolean{blind}}{Anonymous to ISCSLP}
  {
  	$^1$National Taiwan University, Taipei, Taiwan\\
  $^2$NTU Artificial Intelligence Center of Research Excellence (NTU AI-CoRE)
  }
}

\email{
	\ifthenelse{\boolean{blind}}{Anonymous to ISCSLP}
	{b11901154@ntu.edu.tw, hungyilee@ntu.edu.tw}
}

\begin{document}

\thispagestyle{empty}

\maketitle
\begin{abstract}
Large Audio-Language Models (LALMs) excel at audio understanding but expose little about \emph{where} in an audio signal they attend. We introduce \emph{instruction-based vector steering}, which constructs a steering vector by contrasting activations from differently instructed prompts while keeping the audio fixed. Through a systematic probe of LALM attention, we find that---unlike standard prompting or audio-based steering---this intervention significantly redistributes the temporal attention allocated to audio tokens, concentrating it on acoustically relevant regions. We then show that this attention shift is behaviorally meaningful: in a controlled three-event setting, reading out the temporal position of maximal steering-induced attention change recovers the location of a queried sound event without any training, attaining 60.87\% and 68.72\% overlap with ground-truth intervals on Qwen2-Audio and Audio Flamingo 3, far above direct prompting (31.84\%, 46.75\%) and random baselines (27.74\%). Our results characterize a mechanistic property of instruction-based steering in LALMs and provide a training-free probe for the latent temporal structure these models encode.
\end{abstract}


\section{Introduction}

Large Audio-Language Models (LALMs)~\cite{chu2024qwen2audio, goel2025audio, team2025gemma, wu2025step, tang2024salmonn, deshmukh2023pengi} have demonstrated remarkable capabilities in generalized audio understanding and multi-hop reasoning~\cite{yang2025sakura, sakshi2025mmau}. Concurrently, specialized Speech Language Models (SLMs)~\cite{lakhotia2021generative, borsos2023audiolm, wang2023neural, zhang2023speechgpt, defossez2024moshi, fang2024llama} have significantly advanced discrete speech modeling and spoken dialogue. However, both architectures frequently suffer from hallucination~\cite{kuan2024understanding}, generating content ungrounded in the audio input. Recent work in representation engineering~\cite{zou2023representation} shows that activation steering~\cite{zou2023representation, turner2023activation} can mitigate hallucination in multimodal~\cite{li2025vista} and audio models~\cite{lin2025adaptive}. These methods typically rely on \textit{modality-based} contrast, constructing steering vectors by comparing activations with and without the sensory input (e.g., audio vs. silence).

In this work, we introduce \textbf{instruction-based vector steering}. Instead of altering the audio, we contrast activations from differently instructed prompts: one targeting specific acoustic content and another providing generic instructions. Through a systematic probe of LALM attention patterns under various conditions (unsteered, prompt engineering, modality-based steering, and our method), we discover a striking property: \textbf{only instruction-based steering significantly alters the temporal attention distribution over audio tokens}.

This reveals that instruction-based steering uniquely redirects \textit{where} the model attends. To test whether this attention shift is behaviorally meaningful, we read it out as a \textbf{training-free localization probe}: sliding a temporal window across the audio tokens and selecting the position that maximizes the steering-induced attention shift recovers the location of a queried event without any additional training or architectural modification, giving us a window into the model's internal temporal representation. Our main contributions are:
\begin{itemize}
    \item \textbf{Instruction-based vector steering}, a novel inference-time intervention that contrasts differently instructed prompts while preserving the original audio.
    \item A \textbf{systematic attention analysis} revealing that instruction-based steering uniquely alters the temporal attention distribution over audio tokens, unlike standard prompting or modality-based steering.
    \item A \textbf{training-free localization probe} that reads out this attention shift to recover the temporal position of a queried event, providing behavioral evidence that the latent temporal signal is exploitable and characterizing how it varies across positions, layers, and semantic domains.
\end{itemize}

\section{Related Work}

\subsection{Activation Steering in Multimodal Models}
Activation steering controls neural network behavior at inference time without fine-tuning~\cite{zou2023representation}. In text, it has been used to adjust generation style~\cite{konen2024style} and elicit truthfulness~\cite{li2023inference}. Extending to multimodal inputs, VISTA~\cite{li2025vista} and Adaptive Vector Steering~\cite{lin2025adaptive} contrast the presence and absence of sensory inputs (images or audio) to reduce object and acoustic hallucinations. While these methods control \textit{what} information the model generates, our instruction-based steering uses the prompt as the axis of contrast, allowing us to actively manipulate \textit{where} the model focuses its temporal attention.

\subsection{Temporal Localization in Audio Models}
Historically, sound event detection relies on dedicated architectures like AST~\cite{gong2021ast} or PANNs~\cite{kong2020panns}. To grant LALMs similar capabilities, recent approaches explicitly train for temporal grounding: FLAM~\cite{wu2025flam} uses frame-wise contrastive objectives, and TimeAudio~\cite{wang2025timeaudio} fine-tunes LALMs on densely annotated datasets. Despite this, off-the-shelf LALMs still exhibit severe temporal bias and struggle with precise localization~\cite{yao2025notsync}. We take a complementary, interpretability-driven view: rather than training for temporal awareness, we ask whether the information needed to localize an event is already latent in an off-the-shelf LALM's attention, and we use instruction-based steering as a lens to expose it.


\section{Methodology}

We present our approach in three parts. Section~\ref{sec:ibvs} introduces
instruction-based vector steering. Section~\ref{sec:attention} describes
our attention pattern analysis that reveals its unique mechanistic
property. Section~\ref{sec:detector} shows how we read out this property
as a controlled localization probe.

\subsection{Instruction-Based Vector Steering}
\label{sec:ibvs}

Existing audio-based steering methods~\cite{lin2025adaptive} construct
steering vectors by contrasting activations when the sensory input is
present versus absent. We propose an alternative: \textbf{instruction-based
vector steering}, where both instances receive the \textit{same} audio
but differ only in the textual instruction.

Given an audio input $\mathbf{x}_a$ and a user query $\mathbf{x}_q$, we
define a positive instance $X^+ = (\mathbf{x}_a,\; p^+ \oplus
\mathbf{x}_q)$, where $p^+$ is an instruction directing the model to
focus on the meaningful content of the audio (e.g., ``\textit{Focus on
the meaningful part of the audio}''). The negative instance $X^- =
(\mathbf{x}_a,\; p^- \oplus \mathbf{x}_q)$ uses a generic instruction
$p^-$ (e.g., ``\textit{Focus on the entire audio}''). Let $F_l(X)$
denote the hidden state of the residual stream at the final token
position for layer $l$. The steering vector is:
\begin{equation}
    \mathbf{v}^l = F_l(X^+) - F_l(X^-), \quad l \in \{0, \ldots, L-1\}.
\end{equation}
During inference, the steering vector is injected into the residual
stream $\mathbf{h}_{t,l}$ at each generation step $t$:
\begin{equation}
    \tilde{\mathbf{h}}_{t,l} = \mathrm{Norm}\left(\mathbf{h}_{t,l} +
    \lambda\, \mathbf{v}^l\right),
\end{equation}
where $\mathrm{Norm}(\mathbf{u}) = \mathbf{u} \cdot
\frac{\lVert\mathbf{h}_{t,l}\rVert_2}{\lVert\mathbf{u}\rVert_2}$
preserves the original hidden state norm, and $\lambda$ controls the
steering strength.

\subsection{Attention Pattern Analysis}
\label{sec:attention}

To understand \textit{how} different intervention methods affect the
model's internal behavior, we measure the proportion of attention mass
allocated to audio tokens under four conditions: (1)~the original
unsteered model, (2)~prompt engineering (prepending a focus instruction
to the input without steering), (3)~audio-based vector
steering~\cite{lin2025adaptive}, and (4)~our instruction-based vector steering.

For each condition and each layer $l$, we compute the
\textbf{audio attention proportion}:
\begin{equation}
    P_l = \frac{\sum_{i} \sum_{j \in \mathcal{A}} \bar{a}^l_{i,j}}
    {\sum_{i} \sum_{j} \bar{a}^l_{i,j}},
\end{equation}
where $\bar{a}^l_{i,j}$ is the attention weight from position $i$ to
position $j$ at layer $l$ averaged across all heads, and $\mathcal{A}$
denotes the set of audio token positions.

\begin{figure*}[t]
\centering
\includegraphics[width=\textwidth]{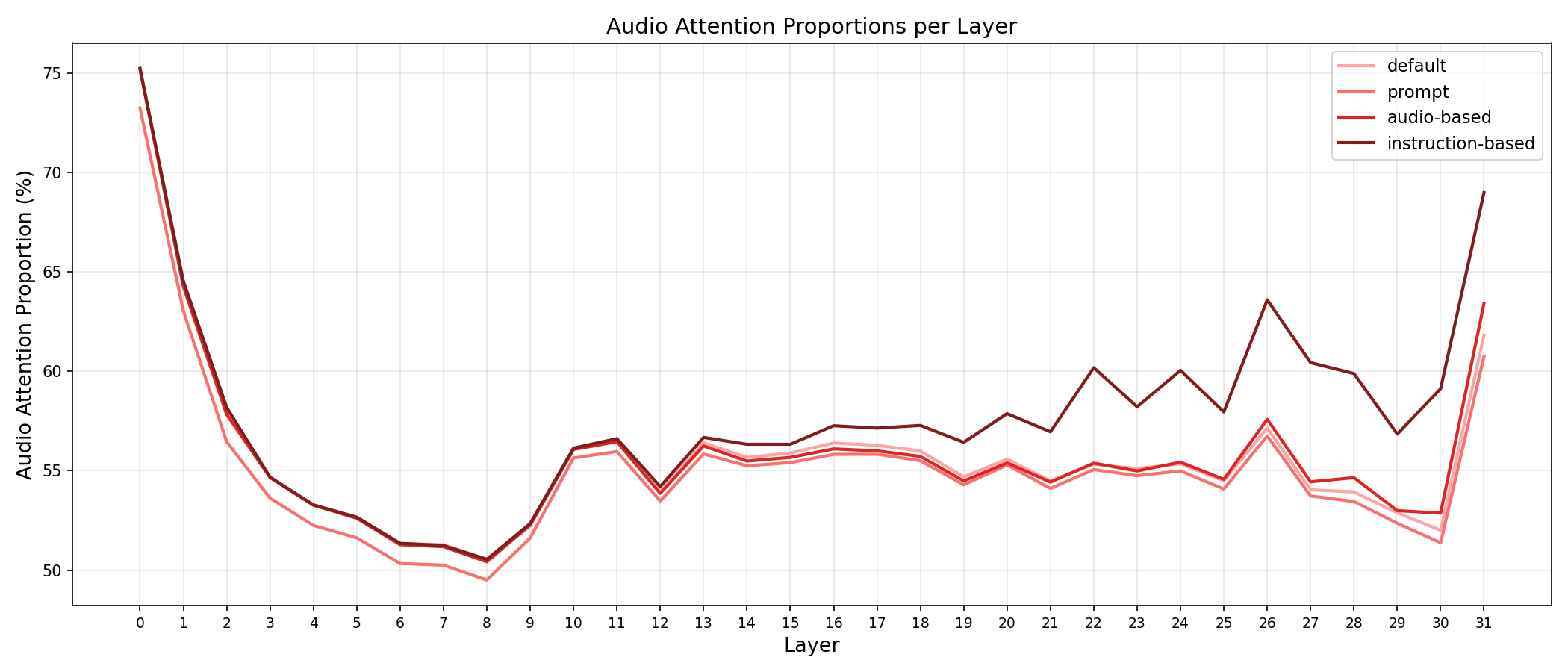}
\caption{Audio attention proportion per layer under four conditions on Qwen2-Audio (MMAU-mini). Only instruction-based steering significantly increases the attention allocated to audio tokens, particularly in later layers.}
\label{fig:attn_proportion}
\end{figure*}

As shown in Figure~\ref{fig:attn_proportion}, only instruction-based steering produces a
significant change in the audio attention proportion, particularly in
the later layers. Prompt engineering and modality-based steering leave
the attention proportions largely unchanged relative to the unsteered
baseline. This indicates that instruction-based steering uniquely
redirects the model's attention toward the audio modality, rather
than merely altering the information content of the hidden states.
This observation motivates our use of the later-half layers in the
localization probe described in Section~\ref{sec:detector}.

\begin{figure*}[t]
\centering
\includegraphics[width=\textwidth]{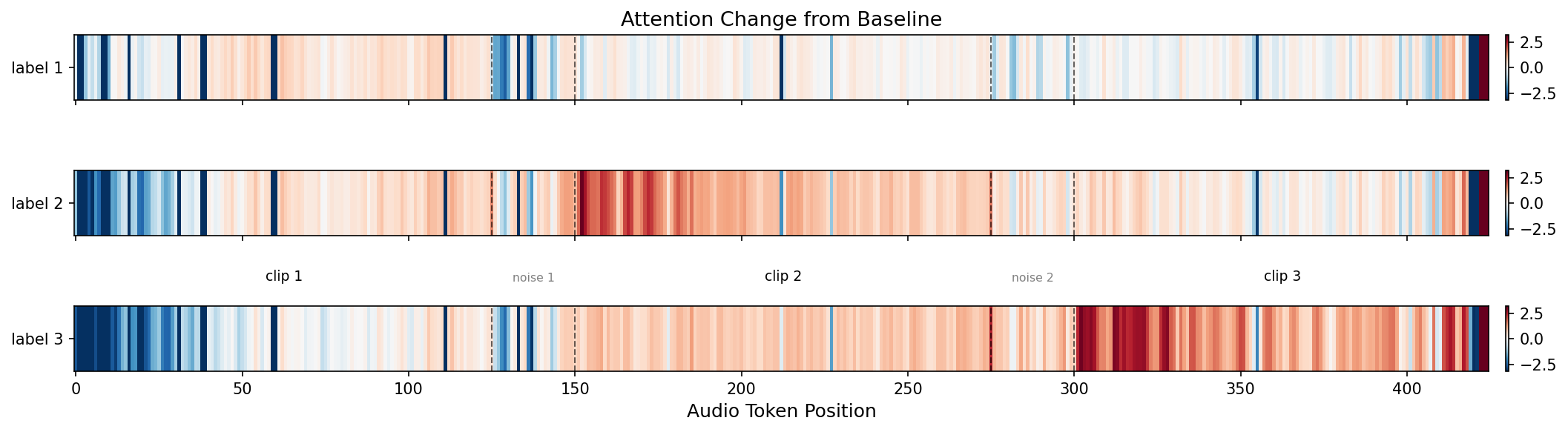}
\caption{Steering-induced attention change per audio token position of Qwen2-Audio, averaged across 48 samples from the controlled benchmark, with each sample containing a sequence of three exactly 5-second clips.  Each row shows the attention delta when steering toward a different label. Attention increases (red) align with the corresponding clip position, confirming that instruction-based steering selectively redirects temporal focus.}
\label{fig:attn_delta}
\end{figure*}
\subsection{Reading Out the Attention Shift as a Localization Probe}
\label{sec:detector}

The finding above implies that instruction-based steering concentrates
attention on acoustically relevant regions. We exploit this property to
read out a training-free localization estimate directly from the model.

Given a composite audio $\mathbf{x}_a$ containing multiple sound events
and a target event description $e$ (e.g., ``\textit{cat}''), we construct
a label-specific steering vector using the positive instruction
$p^+ =$ ``\textit{Which part of the audio has the sound of} $e$\textit{?}''
and the negative instruction $p^- =$ ``\textit{Carefully analyze
everything in this audio.}''

We then perform two forward passes with a generic inference prompt: one
\textbf{unsteered} (base) and one with the label-specific steering vector
\textbf{applied}. From both passes, we extract the attention weights from
the later half of the model's layers ($l \in \{L/2, \ldots, L{-}1\}$),
as these are most affected by instruction-based steering
(Section~\ref{sec:attention}). Since our goal is to localize audio events,
we use only the \textbf{text-to-audio attention}---i.e., the attention
from text token queries to audio token keys---as it most directly reflects
the model's focus on the audio input and avoids the causal bias inherent
in audio-to-audio self-attention.

For each layer $l$, we compute the attention proportion within a sliding
window. Let $\mathbf{a}^l_{\mathrm{base}}$ and
$\mathbf{a}^l_{\mathrm{steer}}$ denote the text-to-audio column sums for
the base and steered passes, respectively. For a window of size $W$
starting at audio token position $i$, the attention proportion under
condition $c \in \{\mathrm{base}, \mathrm{steer}\}$ is:
\begin{equation}
    P^l_c(i) = \frac{\sum_{t=i}^{i+W-1} a^l_{c,t}}
    {\sum_{t} a^l_{c,t}},
\end{equation}
where the denominator sums over all audio token positions. We then compute
the proportion ratio between the steered and base conditions:
\begin{equation}
    R^l(i) = \frac{P^l_{\mathrm{steer}}(i)}{P^l_{\mathrm{base}}(i) + \epsilon},
\end{equation}
where $\epsilon$ is a small constant for numerical stability. The
overall score for each window is the weighted average of $R^l(i)$ across
layers, with linearly increasing weights assigned to later layers.
The predicted event location is:
\begin{equation}
    i^* = \arg\max_i \; \sum_{l=L/2}^{L-1} w_l \, R^l(i),
\end{equation}
where $w_l$ increases linearly from layer $L/2$ to $L{-}1$. The detected
interval $[i^*, \; i^* + W - 1]$ is mapped back to the corresponding time
interval. This probe requires no additional training, fine-tuning, or
architectural modification---it relies solely on the latent temporal
structure revealed by instruction-based steering.


\section{Experiments}

\subsection{Controlled Localization Benchmark}
\label{sec:dataset}

To isolate temporal event localization under controlled conditions, we
construct a benchmark of 500 composed audio samples. Each sample consists
of three clips concatenated sequentially, with each clip lasting
4.5--5.5 seconds. Clips are drawn from two domains: animal sounds (dog,
cat, rooster, cow, frog, crow) and human speech (English, French,
Chinese, Japanese), adapted from the SAKURA dataset~\cite{yang2025sakura}; we use
SAKURA only as a source of cleanly labeled single-event audio, not as an
evaluation benchmark in its own right. For each trial, the three slots
independently sample from the animal or language pool with equal
probability, subject to the constraint that all three clips have distinct
labels. Between consecutive clips, we insert a 1-second noise segment
whose energy is matched to the average per-sample energy across the three
clips, and all clips are normalized to the same average energy.

This controlled composition---sequential, non-overlapping clips with
matched energy and clean boundaries---lets us attribute any change in
temporal attention to the steering intervention itself, rather than to
confounds such as overlap, noise, or boundary ambiguity, which is
essential for the mechanistic analysis that follows.

\subsection{Attention Pattern Analysis Setup}

We evaluate the effect of different intervention methods on audio
attention using the MMAU-mini benchmark~\cite{sakshi2025mmau}. MMAU-mini is used
here solely to measure how each intervention reshapes the layer-wise
audio-attention proportion $P_l$ (Figure~\ref{fig:attn_proportion}); it
is not used for the localization experiments, which rely exclusively on
the controlled benchmark of Section~\ref{sec:dataset}. For each sample,
we run four forward passes under the conditions described in
Section~\ref{sec:attention}: unsteered, prompt engineering, audio-based
steering, and instruction-based steering. We set the steering strength
$\lambda = 0.1$ for both steering methods. We report the audio attention
proportion $P_l$ at each layer, averaged across all samples in the
benchmark.

\subsection{Localization Probe Setup}

We evaluate our localization probe on the controlled benchmark using two
models: Qwen2-Audio-7B-Instruct~\cite{chu2024qwen2audio} (32 layers) and Audio
Flamingo 3~\cite{goel2025audio} (28 layers). For each composed sample, we query
each of the three labels independently and apply the sliding window
method from Section~\ref{sec:detector} using the later half of the
model's layers (layers 16--31 for Qwen2-Audio, layers 14--27 for AF3)
with $\lambda = 0.1$. The window size is set to match the ground-truth
clip length in token space.

We compare against two baselines: (1)~a \textbf{random baseline}, which places the window uniformly at random across valid positions, and (2)~a \textbf{direct prompting baseline}, which provides the model with the exact duration of the target event and asks it to explicitly output the start time (e.g., ``\textit{This audio contains three sound events in sequence. The sound of cat lasts 5.00 seconds. What is the starting time (in seconds) of this event from the beginning of the audio?}"). This predicted start time is then combined with the known duration to construct the estimated window for overlap calculation.

We report the \textbf{overlap percentage}: the proportion of the
ground-truth interval covered by the predicted window, broken down by
target position (begin, middle, end) and overall. Overlap is a natural
readout in our setting, where each sample contains exactly one target
interval of known length.

\subsection{Attention Probing}

To verify that instruction-based steering indeed redirects temporal
attention toward the queried event, we conduct an attention probing
analysis. For each composed sample, we run four forward passes: one
unsteered baseline and three with label-specific steering vectors. We
compute the text-to-audio attention sum at each audio token position,
summed across all layers, and take the difference between each steered
pass and the baseline. Figure~\ref{fig:attn_delta} shows the averaged
attention delta across samples. When steering toward a specific label,
attention increases at the corresponding clip position, confirming that
instruction-based steering selectively redirects temporal focus to the
queried event.

\begin{table}[t]
\centering
\caption{Overlap (\%) between predicted and ground-truth event intervals by target position. \textit{Direct}: directly prompting the model for timestamps. \textit{Window}: instruction-based vector steering for window selection. Best results per column are \textbf{bolded}.}
\label{tab:results}
\setlength{\tabcolsep}{4pt}
\begin{tabular}{llcccc}
\toprule
\textbf{Model} & \textbf{Method} & \textbf{Begin} & \textbf{Middle} & \textbf{End} & \textbf{Overall} \\
\midrule
\multicolumn{2}{l}{Random} & 20.81 & 41.61 & 20.81 & 27.74 \\
\midrule
\multirow{2}{*}{Qwen2} 
  & Direct  & 36.86 & 53.04 & \phantom{0}5.61 & 31.84 \\
  & Window(ours)    & \textbf{36.90} & \textbf{73.39} & \textbf{72.32} & \textbf{60.87} \\
\midrule
\multirow{2}{*}{AF3} 
  & Direct  & \textbf{98.91} & 33.46 & \phantom{0}7.87 & 46.75 \\
  & Window(ours)    & 73.98 & \textbf{48.31} & \textbf{83.87} & \textbf{68.72} \\
\bottomrule
\end{tabular}
\end{table}


\section{Results \& Discussion}

Table~\ref{tab:results} presents the overlap percentages for both models across target positions. 

\textbf{Qwen2-Audio.} The direct prompting baseline achieves only 31.84\% overall, nearly indistinguishable from the random baseline (27.74\%). The per-position breakdown confirms that this is not meaningful localization: performance on End clips drops to just 5.61\%, suggesting the model defaults to early positions when forced to produce timestamps. In contrast, our steering-based probe achieves 60.87\% overall, with strong performance across all three positions (36.90\% / 73.39\% / 72.32\%). This indicates that instruction-based steering exposes a latent temporal localization signal that the model cannot surface through prompting alone.

\textbf{Audio Flamingo 3.} The direct prompting baseline appears stronger at 46.75\% overall, but the positional breakdown reveals a severe bias: 98.91\% on Begin clips versus just 7.87\% on End clips. The model is not localizing events---it is systematically guessing the beginning of the audio regardless of the query. Our probe corrects this bias, achieving 68.72\% overall with a substantially more balanced positional profile (73.98\% / 48.31\% / 83.87\%).

\textbf{Performance by Semantic Domain.} Per-category overlap reveals different sensitivities across architectures. Qwen2-Audio localizes animal sounds robustly, peaking at 73.83\% for frog and 69.61\% for dog, but is weaker on some language segments, notably English (43.97\%) and Chinese (50.46\%). Audio Flamingo 3 is more consistent across domains, excelling at cross-lingual speech (73.84\% French, 71.94\% Chinese) while remaining strong on animal sounds (73.19\% frog, 72.49\% cat). Thus, while instruction-based steering reliably extracts a localization signal, the baseline domain competency of the chosen LALM still bounds zero-shot localization accuracy.

\textbf{Position analysis.} The Middle position is consistently the weakest for our probe on both models. We attribute this to the difficulty of localizing a central segment: boundary clips border only one neighbor, giving a sharper attention contrast, whereas the middle clip is flanked on both sides, diluting the differential. Even so, Middle performance stays well above both baselines.

\textbf{Layer selection ablation.} Using all layers instead of the later half reduces overall overlap by approximately 5\% on both models, confirming that later layers carry the strongest localization signal, consistent with the attention analysis in Section~\ref{sec:attention} and with interpretability findings that later layers handle task-specific reasoning.

\textbf{Attention probing.} Figure~\ref{fig:attn_delta} shows why the sliding window probe works: when steering toward a specific label, the attention delta concentrates at the corresponding clip position while attention elsewhere decreases. The effect manifests differently across positions. For the first segment (top row), the absolute increase (red) is less dramatic than the spikes for later clips, but steering toward it sharply reduces the attention loss (deep blue) that the first segment otherwise suffers when steering toward the second or third labels (middle and bottom rows)---acting as an attention preserver rather than amplifier. This pattern is consistent across all three positions and robust when averaged over 48 samples. The shift is sharpest at segment boundaries, matching our finding that Begin and End clips are easier to localize than Middle clips.

\textbf{Cross-model consistency.} Our steering-based probe produces substantial improvements on both Qwen2-Audio and Audio Flamingo 3, demonstrating its effectiveness as a general technique for different LALMs. 

\section{Conclusion}

We introduced instruction-based vector steering for Large Audio-Language Models and discovered that it uniquely alters temporal attention patterns over audio tokens---a property not shared by prompt engineering or audio-based steering. Through attention analysis, we showed that this effect is concentrated in the later layers and manifests as a selective redistribution of attention toward acoustically relevant regions. Exploiting this finding, we proposed a training-free localization probe that reads out the steering-induced attention shift to recover the temporal position of a queried event. Our probe achieves 60.87\% and 68.72\% overlap on Qwen2-Audio and Audio Flamingo 3, substantially outperforming direct prompting baselines that either fail entirely or exhibit severe positional bias. These results demonstrate that pretrained LALMs encode richer temporal information than their text-generation interface reveals, and that activation steering provides a principled tool for accessing this latent structure without additional training.

\bibliographystyle{IEEEtran}

\bibliography{mybib}


\end{document}